\documentclass[letterpaper, 10 pt, conference]{ieeeconf}  

\IEEEoverridecommandlockouts                

\usepackage{comment}
\usepackage{color}
\usepackage{graphicx}
\usepackage[space]{grffile}
\usepackage{multirow}

\usepackage{times} 
\usepackage{amsmath} 
\usepackage[linesnumbered,ruled]{algorithm2e}
\usepackage{subcaption}
\usepackage[export]{adjustbox}
\usepackage[english]{babel}
\usepackage[utf8]{inputenc}
\usepackage[noend]{algpseudocode}
\usepackage{caption}

\usepackage{xcolor}
\usepackage{soul}
\usepackage{bm}
\usepackage[utf8]{inputenc}
\usepackage{array}
\newcolumntype{M}[1]{>{\centering\arraybackslash}m{#1}}
\DeclareMathOperator*{\argmin}{arg\,min}
\let\labelindent\relax
\usepackage[inline]{enumitem}
\usepackage{graphicx}

\title{\LARGE \bf
Traffic Management Strategies for Multi-Robotic Rigid Payload Transport Systems
}

\author{Yahnit Sirineni, Pulkit Verma, Kamalakar Karlapalem
\thanks{Yahnit Sirineni and Pulkit Verma are students of Agents and Applied Robotics Group(AARG), Kohli Center on Intelligent Systems(KCIS), International Institute of Information Technology, Hyderabad (IIIT-H), India
       {\tt\small sirineni.yahnit@research.iiit.ac.in}}%
\thanks{Kamalakar Karlapalem is a faculty of Computer Science with International Institute of Information Technology, Hyderabad, India
       {\tt\small kamal@.iiit.ac.in}}
}

\begin{document}
\maketitle
\thispagestyle{empty}
\pagestyle{empty}

\begin{abstract}
In this work, we address traffic management of multiple payload transport systems comprising of non-holonomic robots. We consider loosely coupled rigid robot formations carrying a payload from one place to another. Each payload transport system (PTS) moves in various kinds of environments with obstacles. We ensure each PTS completes its given task by avoiding collisions with other payload systems and obstacles as well. Each PTS has one leader and multiple followers and the followers maintain a desired distance and angle with respect to the leader using a decentralized leader-follower control architecture while moving in the traffic.  We showcase, through simulations the time taken by each PTS to traverse its respective trajectory with and without other PTS and obstacles. We show that our strategies help manage the traffic for a large number of PTS moving from one place to another.
\end{abstract}

\section{INTRODUCTION}
Multi-robot systems are a well studied problem in robotics where applications like payload transportation \cite{pulkit_arms}\cite{transport1}\cite{transport2}, traffic management \cite{traffic1}\cite{traffic2}, and area exploration \cite{ae1}\cite{ae2} are most commonly used. A system with multiple robots/agents offers several advantages like low power consumption, increased redundancy, and makes the system modular \cite{cao}\cite{parker}\cite{Yang}. Handling more number of robots poses different challenges on the functionality of the system. Collision avoidance of a system with other neighbourhood transport systems and other obstacles in the environment is an important challenge. The basic idea of collision avoidance is that a system derives the control inputs based on its surroundings such that it does not collide with any other system or obstacles while performing its task. Work has been carried out on avoiding the obstacles for a single or multiple robots \cite{orca}, further research is needed on obstacle avoidance for multiple robots moving in a formation(s) of different shapes.\par   
For payload carrying tasks in industries, assembly lines, warehouses, etc., where payloads need to be transported from one place to another without affecting the functionality of other systems, is a challenging problem. Such applications require smooth and collision free movement of these systems such that \begin{itemize}
    \item A PTS does not collide with the other payload systems in the surroundings.
    \item Avoids collisions with static obstacles.
    \item Reaches the destination by following the desired trajectory.
\end{itemize}
In our work, we simulate multiple PTS carrying payloads of different shapes (like triangular, square, circle, etc.) from one point to another. Each system comprises of a group of non-holonomic robots moving in a formation using a decentralized leader-follower approach in \cite{formation}. One leader and multiple followers are present in each system. The leader uses the trajectory information provided by our control architecture which is derived from source to destination. The followers generate their velocities using the leader information which makes the formation intact all the time. We show, through simulations, the time taken by a PTS to move from one point to another in an environment with \begin {enumerate*}[label=(a),font=\itshape]
    \item Only one payload transport system.  
\end{enumerate*} 
\begin{enumerate*}[label=(b),font=\itshape]
    \item Multiple PTS with a possibility of having a collision 
\end{enumerate*} 
\begin{enumerate*}[label=(c),font=\itshape]
    \item Multiple PTS and static obstacles in the environment.
\end{enumerate*}
\par
We propose a modified version of Optimal Reciprocal Collision Avoidance for non-holonomic robots (nh-ORCA) \cite{nh-orca} algorithm to identify if a PTS is prone to collide with other formations and obstacles in the environment. Our method provides the command velocities to the robots in the formation such that no collision happens while the transport system is approaching its goal and also ensures least deviation possible to the existing path of the formation. We present the simulation results for different formation shapes with and without obstacles. \par
We discuss the related work done in this domain in Section \ref{related}. Our contribution in this paper is discussed in Section \ref{contribution}. The modeling and control of non-holonomic robots is carried out in Section \ref{Modeling}. Traffic Management strategies for multiple PTS are discussed in Section \ref{ca}.  To make our solution more concrete, we display simulation results in Section \ref{results}. Conclusion is discussed in Section \ref{conclusion}


\section{Related Work}
\label{related}
Over the years, various methods have been developed for formation control and navigation of formations for payload transportation. \cite{liu2011coordinated}, \cite{pereyra2017path} propose methods for path planning of a single formation without laying much emphasis on the presence of obstacles.  \cite{shapira2015path},  \cite{polkowski2010navigation}, \cite{garrido2013general},    \cite{gautam2018motion} shows navigation of single formation in environments which consists of static obstacles.Path planning for 3D formations is proposed in \cite{alvarez20153d} and \cite{zuo2016new}.  Method for navigating a single formation using the leader-follower approach is shown in \cite{garrido2013general} and \cite{asl2014control}. \cite{alonso2016distributed},\cite{zhou2018agile} shows some promising results in aerial vehicle formation control. A slung payload transportation method is demonstrated in \cite{bernard2009generic} where as \cite{hou2008wheeled} devised wheeled locomotion for payload carrying with modular robot. \cite{alonso2017multi} proposes path planning of a single formation in environment with dynamic obstacles but is a computationally expensive centralized method. 

To the best of our knowledge, no work has been done on path planning and collision avoidance of multiple formations of robots in environments with obstacles. We cannot consider other formations as dynamic obstacles and use the approach presented in \cite{alonso2017multi} because the formations have a specific goal of reaching the destination while dynamic obstacles do not have a proper goal and hence show random behaviour. This aspect makes our task even more challenging .

\section{Contribution}
\label{contribution}
We propose a Leader-Follower-ORCA-RRT* framework which is a decentralized method for navigation of multiple formations of robots carrying payloads to the desired destination while avoiding collisions with other formations and obstacles. Given an environment in which multiple payloads are to be transported from one place to another through formation of robots, we first compute the path of the leader of the formation using RRT* \cite{rrt} from respective source to destination considering the static obstacles and interpolate the path to get multiple way points. We modify nh-ORCA \cite{nh-orca} to compute  collision-free linear and angular velocity for the leader of the formation. A leader-follower based decentralized control law is incorporated to compute the corresponding command velocities of the followers in the formation. Finally, We present simulation results of the implemented control architecture to validate our method. 
\par

\section{Background}
\label{Modeling}
\subsection{Non-holonomic Robot Kinematics}
Consider a non-holonomic mobile robot positioned at $(x, y)$ and oriented at and angle $\theta$. Assuming the pseudo-command velocities be $v, \omega$, the robot kinematics is given as
\[
  \left[ {\begin{array}{c}
   \dot{x}\\
   \dot{y}\\
   \dot{\theta}\\
  \end{array} } \right] =
  \left[ {\begin{array}{cc}
   cos\theta & -d sin\theta\\
   sin\theta & d cos\theta\\
   0 & 1\\
  \end{array} } \right] \left[ {\begin{array}{c}
   v\\
   \omega\\
  \end{array} } \right]
\]
where $d$ is the distance of the robot's center to the center of mass of the robot.
\subsection{Leader-Follower Formation Control} 
\begin{figure}[h]
    \centering
	\includegraphics[width=\linewidth, height=3cm]{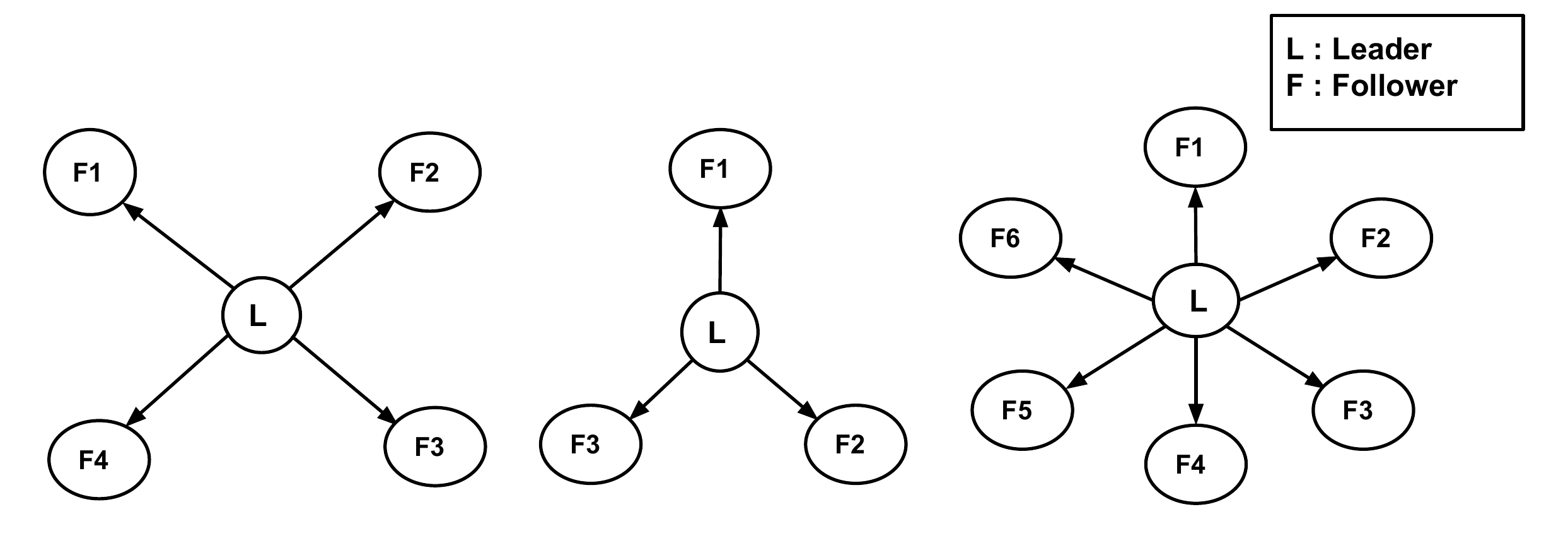}
	\caption{Leader follower formation with different shapes}
	\label{formation_shapes}
\end{figure}

Each PTS comprises of a number of non-holonomic wheeled mobile robots which move on a trajectory while maintaining a rigid formation. A decentralized leader-follower approach control algorithm \cite{formation} is used which include a leader and multiple followers. The followers use the command velocities and odometery information of the leader to derive their velocities such that the followers maintain a specified distance and angle to remain in a formation. Fig. \ref{formation_shapes} shows different shapes of the formation used in the paper. The formation control law is given by the following equations
\begin{eqnarray} 
\label{formation_control_eqn}
v_{j} &=& (k_{1} \alpha_{j} + v_{i} cos \theta_{ij} - \rho^{d}_{ij} \omega_{i} sin(\psi^{d}_{ij} - \theta_{ij}) \\
\omega_{j} &=& \frac{v_{i}sin\theta_{ij} + \rho^{d}_{ij} \omega_{i} cos(\psi^{d}_{ij} + \theta_{ij}) + k_{2} \beta_{j} + k_{3}\theta_{je}}{d} \nonumber
\end{eqnarray}
where $\alpha_{j}$ and $\beta_{j}$ is the error in longitudinal and vertical direction respectively. Constants $k_{1}, k_{2}, k_{3}, k_{4}, k_{5}, k_{6} > 0$, $\rho^{d}_{ij}$ and $\psi^{d}_{ij}$ are the desired distance and orientation to maintain between the leader and follower robot, $v_{i}$ and $\omega_{i}$ are command velocities of the leader, $v_{j}$ and $\omega_{j}$ are the generated command velocities of the $j^{th}$ follower, $d$ is distance from the robot's center to the robot's center of mass. $\theta_{ij}$ is the orientation error and $x_{je}, y_{je}, \theta_{je}$ are the positional tracking errors between the leader and follower.\par

\subsection{Optimal Reciprocal Collision Avoidance for non-holonomic robots (nh-ORCA)}
nh-ORCA \cite{nh-orca} is a collision avoidance algorithm for non-holonomic robots which is an extension to ORCA\cite{orca} which deals with only holonomic robots.
Using the algorithm presented in \cite{nh-orca}, each robot independently computes its velocity such that it is collision free for at least $\tau$ time assuming that the other robots also compute their velocity using the same method. There is no explicit communication amongst the robots to choose their velocities, hence making this is a robust decentralized system.  Each robot $i$ constructs $ORCA^{\tau }_{i|j}$ $\forall \enspace j \neq i$  which in turn is computed using velocity obstacles presented in \cite{vel_obstacle}.  

Let $p_i$, $p_j$ be the positions of two robots $i$ and $j$ and $r_i$, $r_j$ be the radius of the robots respectively,
$VO^{\tau }_{i|j}$ is the set of velocities of robot $i$ w.r.t robot $j$ such that:
 \begin{equation}
 \label{vo}
     VO^{\tau }_{i|j} = \{ \bar{\textbf{v}} \mid \exists t \in \left [ 0, \tau \right ], t \cdot \bar{\textbf{v}} \in D \left ( p_{j} - p_{i}, r_{i} + r{j} \right ) \}
 \end{equation}
where 
\begin{equation}
    D\left ( p, r \right ) = \left \{ q \mid \parallel q-p \parallel < r \right \} \nonumber
\end{equation}
 
Then $ORCA^{\tau }_{i|j}$ is computed from Eq (\ref{orca_1}):
\begin{equation}
\label{orca_1}
    ORCA^{\tau }_{i|j} = \left \{ v_{H_{i}} \mid \left ( v_{H_{i}} - \left ( v_{i}^{opt} + 0.5*u \right ) \right ).\textbf{n} \ge 0 \right \}
\end{equation}

where $n$ denotes the outward normal of the boundary of $VO_{i|j}$ at ($v_{i}^{opt}$-$v_{j}^{opt}$) + $u$, $u$ is computed as shown in  \cite{orca}.

Now, the set of collision free velocities for robot $i$ is given by Eq (\ref{orca_2}).
\begin{equation}
\label{orca_2}
    ORCA^{\tau }_{i} = S_{AHV_{i}} \cap \bigcap_{j \neq i} ORCA^{\tau }_{i \mid j}
\end{equation}
with $S_{AHV_i}$ being the set of allowed velocities.
The final optimal velocity of the robot is given by equation (\ref{orca_3}).
\begin{equation}
\label{orca_3}
    \textbf{v}_{H_{i}}^{*} = \argmin_{\textbf{v}_{H_{i}} \in ORCA_{i}^{\tau}} \parallel \textbf{v}_{H_{i}} - \textbf{v}_{i}^{pref}
    \parallel
\end{equation}

We extend this approach to our problem for avoiding collisions among other formations and obstacles.
\section{Traffic Management}
\label{ca}
\begin{figure}[h]
    \centering
	\includegraphics[width=\linewidth, height=4.5cm]{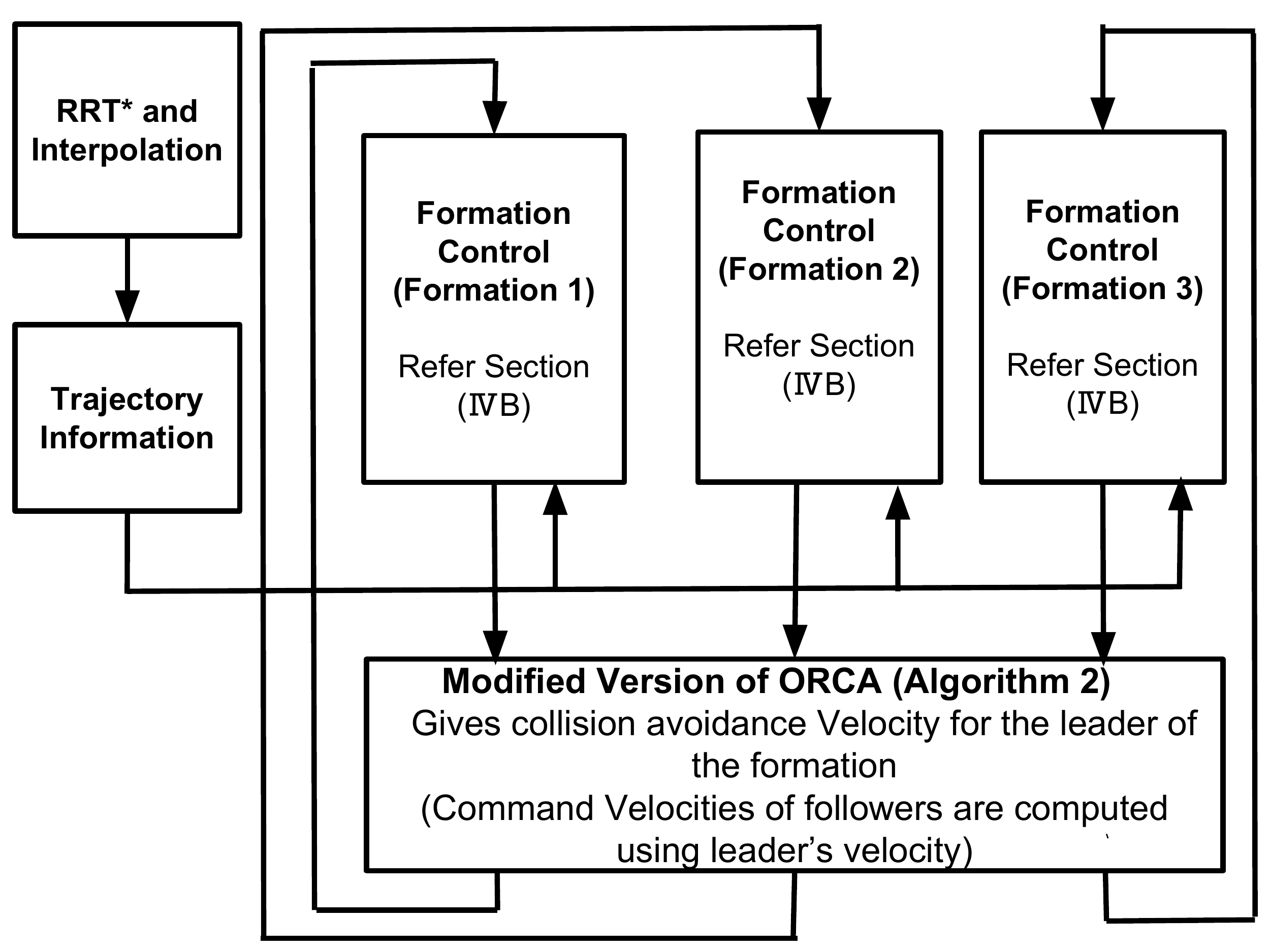}
	\caption{Architecture Diagram of the framework}
	\label{arch}
\end{figure}

Given an environment in which multiple payloads are to be transported from one place to another with multi-robot payload transport systems, we propose an efficient traffic management strategies for path planning and collision avoidance of formations carrying payloads. A bio inspired neurodynamic based leader follower-ORCA-RRT* framework is discussed, which computes a path to the respective destination for all the leaders of PTS considering the static obstacles using RRT* \cite{rrt}, then uses a modified version of nh-ORCA \cite{nh-orca} for collision avoidance amongst other formations and finally incorporates a bio inspired neurodynamic based leader-follower formation control law for computing the command velocities for the followers in the formation. The architecture diagram has been shown in Fig.\ref{arch}
\begin{algorithm}
\caption{Leader Follower-ORCA-RRT*}
\label{alg:algo1}
\SetAlgoLined
\textbf{Input} : environment representing space and obstacles, source and destination of the formations, Configuration of each formation \;
\For{$i=1, i \leq No. \enspace of \enspace Formations, i++$}
{
     \tcp{Compute path from src to dest for $i$th formation using RRT*}
     $path_i$ = RRT*($source_i$, $destination_i$) \;
     
    \tcp{Interpolate the path returned}
      $path_i$ = Interpolate($path_i$) \;
    
    \tcp{Initialize $nextDest_i$ of formation}
     $nextDest_i$ = $path_i$[1]
     
     \tcp{Initialize $v^{pref}_{i}$ of formation}
     $v^{pref}_{i}$ = ($nextDest_i$ - $p_i$)

}
\While{All Formations have'nt reached their destinations}
{
    \For{$i=1, i \leq No. \enspace of \enspace Formations, i++$}
    {
        $timestep$ ++;\;
     \uIf{$f_i$ has not yet reached}
        {
            \tcp{Compute command velocities of the leader}
            $v_{i}, \omega_{i}$ = Algorithm2(i)\;
            \tcp{Compute command velocites of the followers using leader-follower control law}
             LeaderFollower($v_{i}, \omega_{i}$)\;
            Apply corresponding controls to the leader and followers of the formation\; 
        }
    }
}
\end{algorithm}

\begin{algorithm}
\caption{Modified nh-ORCA}
\label{alg:algo2}
\SetAlgoLined
\textbf{Input} : $path_i$, $p_i$, $v_i$, $r_i$, $v^{pref}_{i}$, $v^{max}_{i}$, $neighbours_i$\;
\For{$j \enspace in \enspace neighbours_i$}
{
    Compute $VO^{\tau}_{i}$ from Eq \ref{vo}\;
    Compute $u$ from \cite{orca}\;
    Compute $ORCA^{\tau }_{i|j}$ from Eq \ref{orca_1}
}
Compute $ORCA^{\tau}_{i}$ from Eq \ref{orca_2}\;
Compute the optimal holonomic velocity of the leader $v^{*}_{H_i}$ from Eq(\ref{orca_3})\;
Compute the optimal linear and angular velocity of the leader from $v^{*}_{H_i}$ for non-holonomic robot using nh-ORCA \cite{nh-orca}\;
\uIf{distance($p_i$, $nextDest_i$) $\leq \delta$}
{
    \tcp{If the formation has almost reached its $nextDest$, then update $nextDest_i$ and $v^{pref}_{i}$}
    Update $nextDest_i$ to next point of interpolated path\;
    $v^{pref}_{i}$ =  ( $nextDest_i$ - $p_i$ )\;
}

\textbf{return} non-holonomic command velocities of the Leader.
\end{algorithm}

Multiple PTS are considered in this work. Each PTS is a set of robots with a leader and its followers working together to perform a particular task. Follower $j$ of each formation ($f_i$) maintains a certain angle and distance relative to its leader.  
Each formation $f_i$ has a current position $p_i$, radius $r_i$,  current velocity $v_i$, number of followers $followers_i$, source $src_i$, destination $dest_i$, next immediate destination $nextDest_i$, preferred velocity $v^{pref}_{i}$ i.e the velocity with which it would traverse had there been no obstacles in its path,  maximum velocity $v^{max}_{i}$, neighbours $neighbours_i$ which indicates the formations in its vicinity, $neighbourDist_i$ which refers to the formations that are at a distance less than this value.\par
We use the RRT* to find the preferred path from $src_i$ to $dest_i$ for each formation $f_i$ and interpolate the path returned by RRT* to get multiple way points in the path. The $nextDest_i$ for each formation $i$ is set to the first point of the interpolated path indicating that the formation's tentative target is to reach the $nextDest_i$ and $v^{pref}_{i}$ for each formation is set to ($nextDest_i$ - $p_i$).\par
For each timestep $dt$, new collision avoiding velocities are computed for each of the leaders of the formations through Algorithm \ref{alg:algo2}. To compute the collision avoidance velocities for each formation $f_i$, consider all the neighbours $j$ in $neighbours_i$ for formation $f_i$ and compute $VO^{\tau }_{i|j}$ and $ORCA^{\tau }_{i|j}$ from Eq.(\ref{vo}) and Eq(\ref{orca_1}) for all the neighbours $j$ in $neighbours_i$ and compute $u$ from \cite{orca}. Now, compute $ORCA^{\tau }_{i}$ and $\textbf{v}_{H_{i}}^{*}$ from Eq(\ref{orca_2}) and Eq(\ref{orca_3}) to get the holonomic velocity of the leader of the formation $f_i$. Now, map the holonomic velocity to non-holonomic velocities as illustrated in \cite{nh-orca}. If the formation $f_i$ at a distance less than $\delta$ from $nextDest_i$ where $\delta$ is a parameter to be set based on the environment, then update the $nextDest_i$ to the next point of the interpolated path and the preferred velocity of formation $v^{pref}_{i}$ to ($nextDest_i$ - $p_i$). This modified version of nh-ORCA is given in Algorithm \ref{alg:algo2}.\par
Once the leader velocities of the formations are computed using Eq.(\ref{alg:algo2}), command velocities of all the followers are computed using Eq. (\ref{formation_control_eqn}) which are then applied to all the robots. The complete algorithm is illustrated in Algorithm \ref{alg:algo1}
\section{Results}
\vspace{-0.2cm}
\label{results}
\begin{table}[h]
\centering
\captionsetup{ width= 65mm}
\scriptsize
\captionof{table}{List of parameters for each robot}
\begin{tabular}{ |M{4cm}||M{2cm}|  }
 \hline
 Parameter Name & Value\\
 \hline
 $k_{1}$ & 1.5\\
 $k_{2}$ & 1.0\\
 $k_{3}$ & 0.025\\
 $k_{4}$ & 15.0\\
 $k_{5}$ & 1.0\\
 $k_{6}$ & 1.0\\
  $\delta$ & 1.3\\
 $\tau$ & 11\\
  $dt$ & 0.0167\\
  $\tau_{obstacle}$ & 5\\
 $maxspeed$ (m/s) & 0.03\\
 $neighbourDist$ & 4\\
 \hline
\end{tabular}
\label{param}
\normalsize
\end{table}

We showcase the simulation results of the presented algorithm and discuss various aspects of the system. In the simulations,we consider a \textit{Baseline} scenario in which a payload transport system is carrying a payload from one place to another while there are no other transport systems and  static obstacles in the surroundings. We compare the baseline results with \begin{enumerate*}[label=(a),font=\itshape]
    \item Multiple PTS moving in the environment
\end{enumerate*}  
\begin{enumerate*}[label=(b),font=\itshape]
    \item Multiple PTS and static obstacles in the environment
\end{enumerate*}. We compare the time taken, velocity variations and scalability aspect of this approach with the baseline result. 
\subsection{Multiple Payload Transport Systems}
In this case, we consider four payload transport systems carrying a payload to their respective destinations. Each PTS has to avoid the traffic as posed by the other transport systems while heading towards the destination simultaneously. 
\begin{figure}[h]
    \centering
	\includegraphics[width=\linewidth, height=5cm]{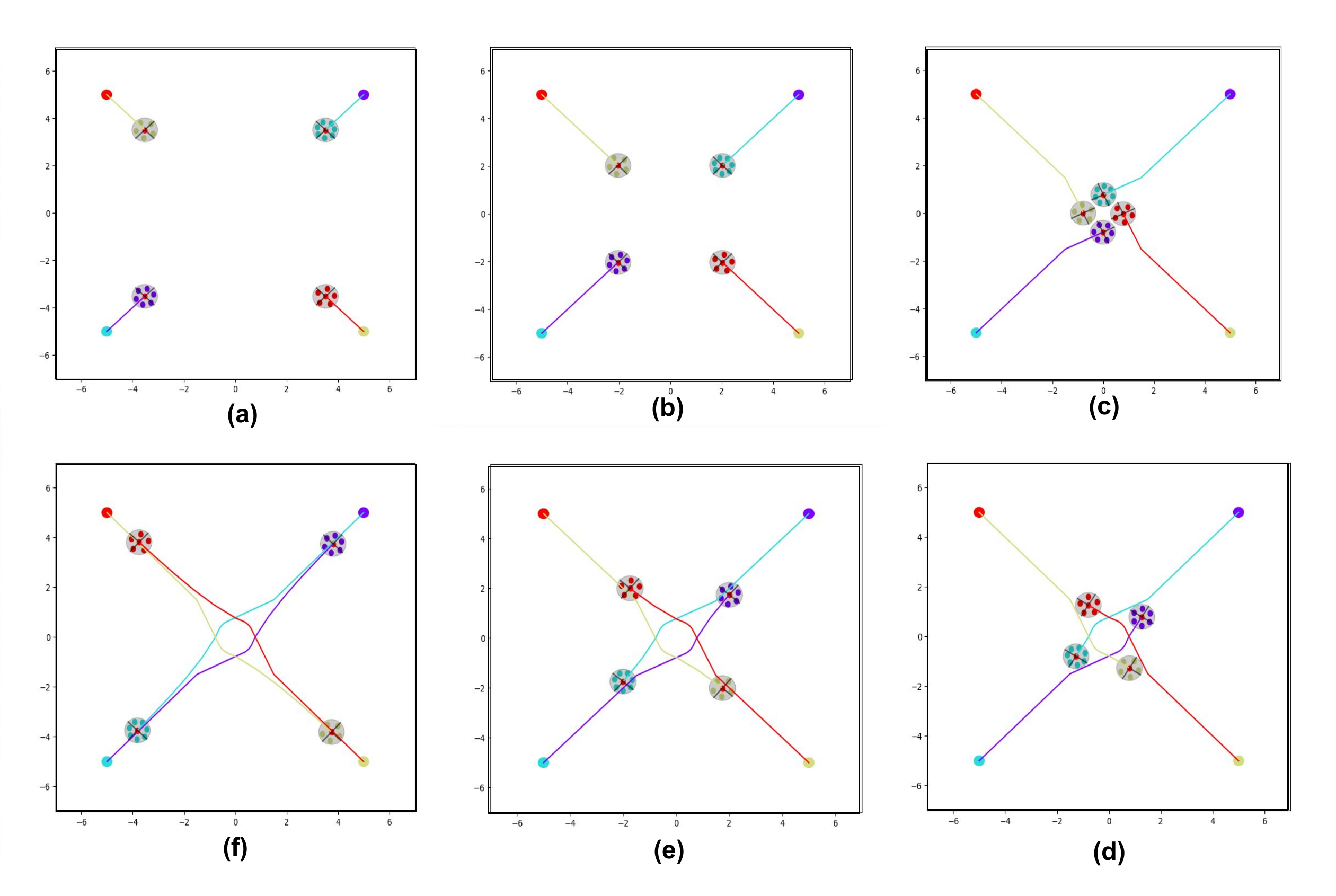}
	\caption{Four PTS are running avoiding collision with the other systems and simultaneously reaching towards the goal}
	\label{no_obs_setup}
\end{figure}
Different shapes (circle, square, triangle) are considered for different PTS to make the problem more challenging. Each formation has different number of robots ranging from $3-10$. However, the number of formations in an environment and number of robot in a formation is not limited and will discussed in latter part of the section. Fig. \ref{no_obs_setup} shows the initial set of all the Payload transport systems present and their respective goal points (denoted by the same color). The simulation is symmetric and hence all formations travel almost the same distance.\par
\begin{figure}[h]
    \centering
	\includegraphics[width=\linewidth, height=4.5cm]{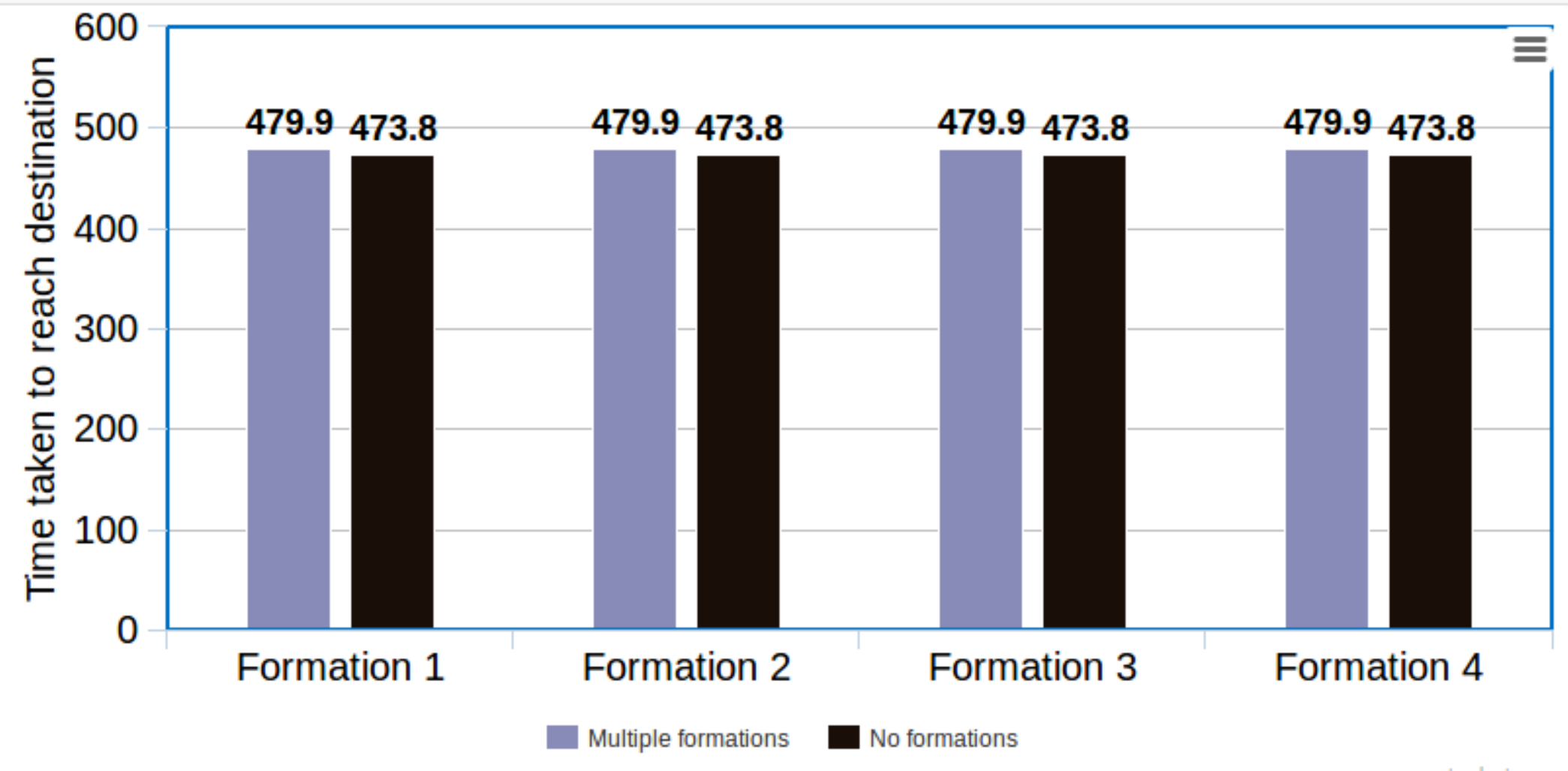}
	\caption{Comparison of time taken by our system with and without any traffic in the environment}
	\label{no_obs_time}
\end{figure}
We compare the time taken by the system in this environment with a system having no traffic in the environment. Fig. \ref{no_obs_time} shows that the time taken by the system in presence of traffic is approximately equal to a system moving in no traffic environment. This time is nearly equal, as all the PTS create a symmetric trajectory and hence takes nearly equal time to reach their destination. \par
\begin{figure}[h]
    \centering
	\includegraphics[width=\linewidth, height=5cm]{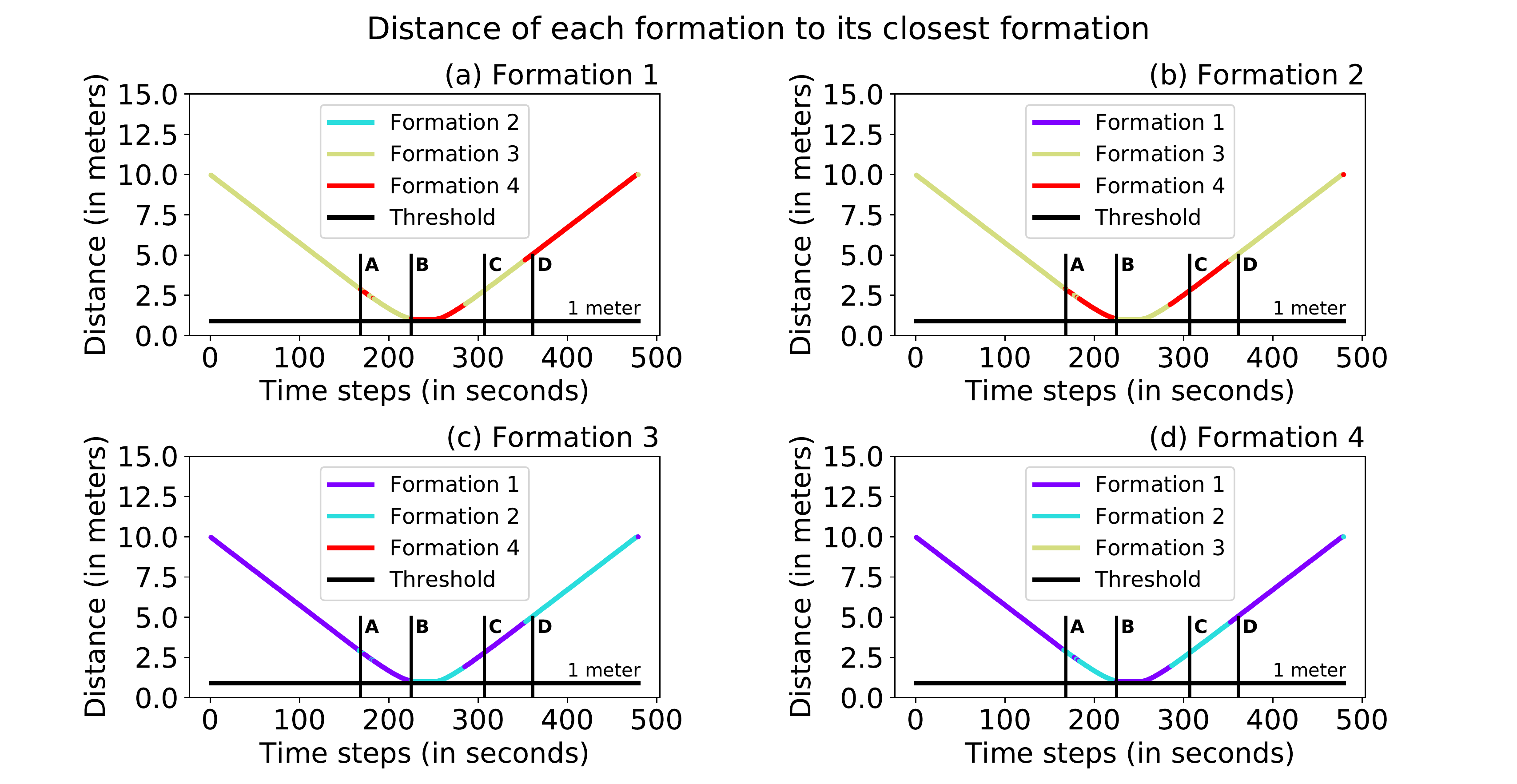}
	\caption{Distance of each PTS with its closest neighbourhood.}
	\label{no_obs_dist}
\end{figure}
\begin{figure}[h]
    \centering
	\includegraphics[width=\linewidth, height=5cm]{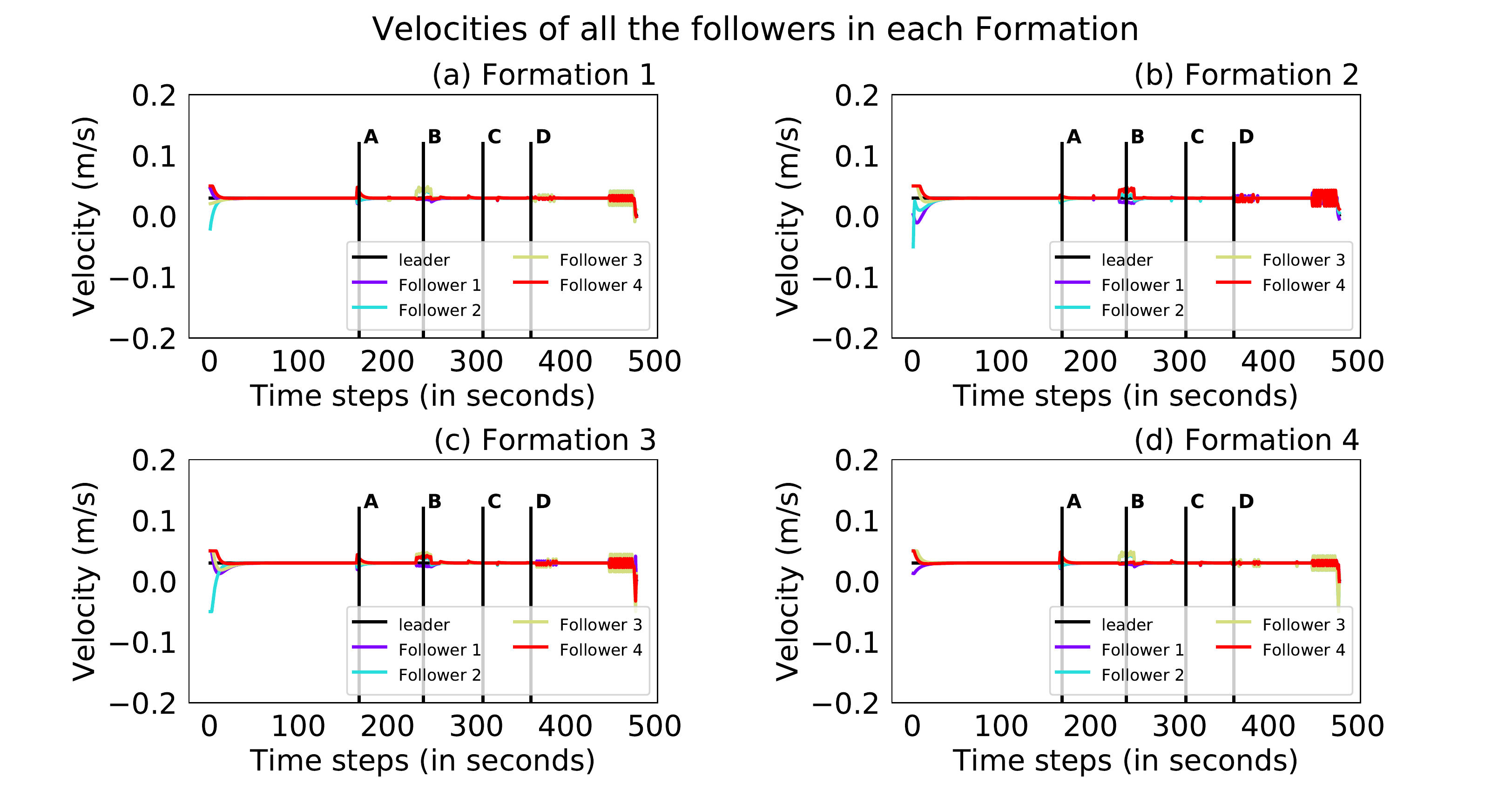}
	\caption{Velocities of all the robots in each PTS.}
	\label{no_obs_vel}
\end{figure}

It is important to note that all the payload transport systems maintain a minimum distance from each other ensuring no collision while in motion. Fig. \ref{no_obs_dist} shows the distance maintained by each PTS to its closest neighbourhood. The figure shows that all the PTS are maintaining a minimum threshold and hence it is clear that the payload systems are not colliding with each other. The change in the velocities of each PTS when coming close to each other is shown in Fig. \ref{no_obs_vel}. We can see that whenever the PTS are coming close to each other, the velocities of the followers in the formation experiences a changes such that the collision can be avoided. Points A, B, C, D are marked in Fig. \ref{no_obs_dist} and Fig. \ref{no_obs_vel} which can be used to relate the two graphs. We can observe that at point B, all the formation are close to each other (see Fig. \ref{no_obs_setup}(f)) and hence suffers a velocity change in the robots of all the PTS. Similarly, point A, C are the points where the PTS started changing the trajectory to successfully avoid the obstacles. \par
\begin{figure}[h]
    \centering
	\includegraphics[width=\linewidth, height=5cm]{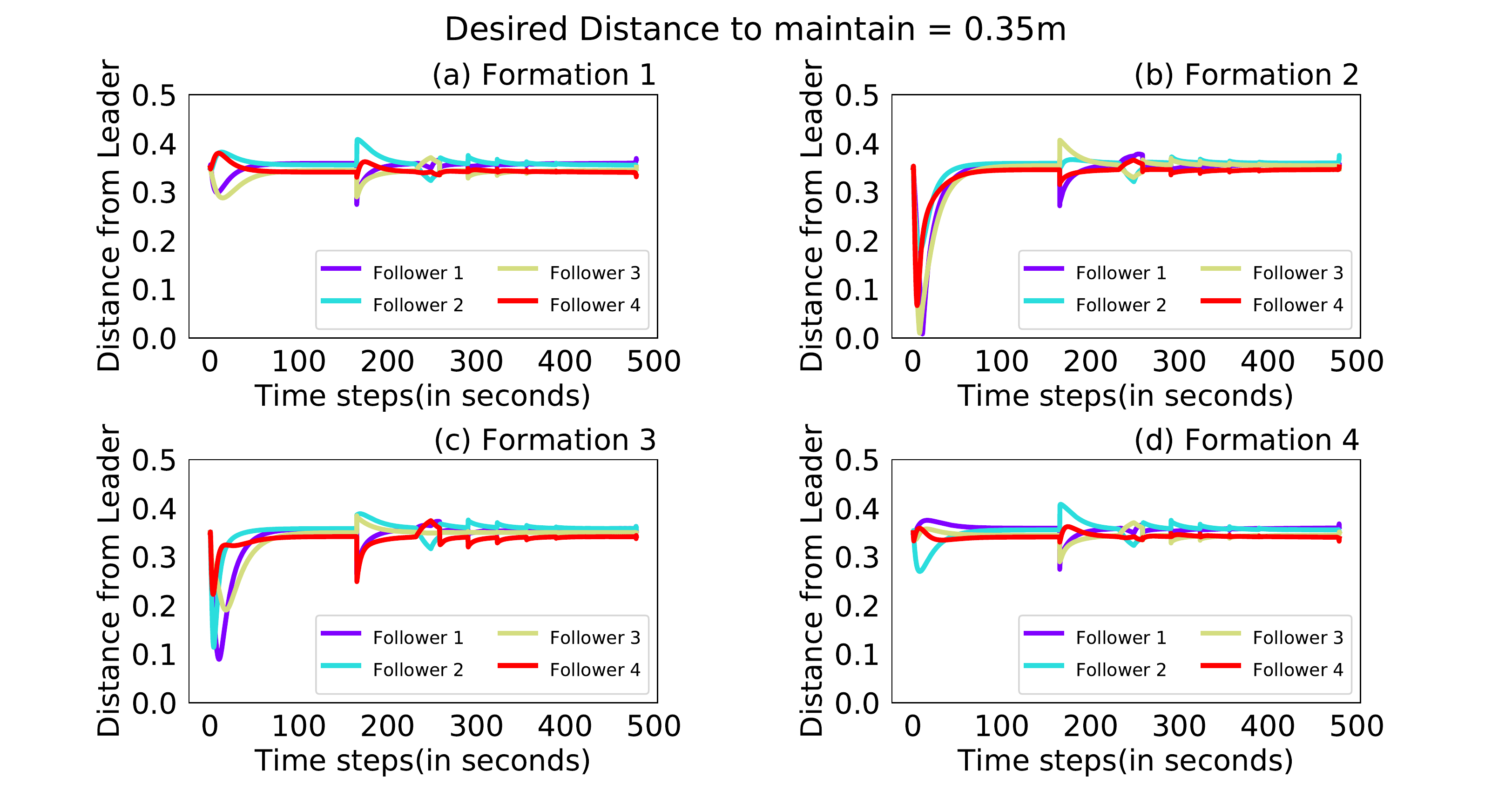}
	\caption{Distance of all the robots in each formation}
	\label{no_obs_fd}
\end{figure}
While avoiding traffic in the surroundings, we ensure that each payload transport system maintains the robots formation using the leader-follower based approach. Fig. \ref{no_obs_fd} depicts that the followers in each formation nearly maintains the desired distance and orientation from the leader. For simulation, we have considered this distance constant for all the followers in all the PTS, which is equal to $0.35 m$. The figure shows that the distance of each followers from the leader is approximately equal to $0.35$, even when a system avoids any collision with its neighbourhood.
\subsection{Multiple Payload Transport Systems with Static Obstacles}
To make our results more concrete, we consider an environment where there are multiple PTS are moving along with the static obstacles between the start point and the destination. 
\begin{figure}[h]
    \centering
	\includegraphics[width=\linewidth, height=6cm]{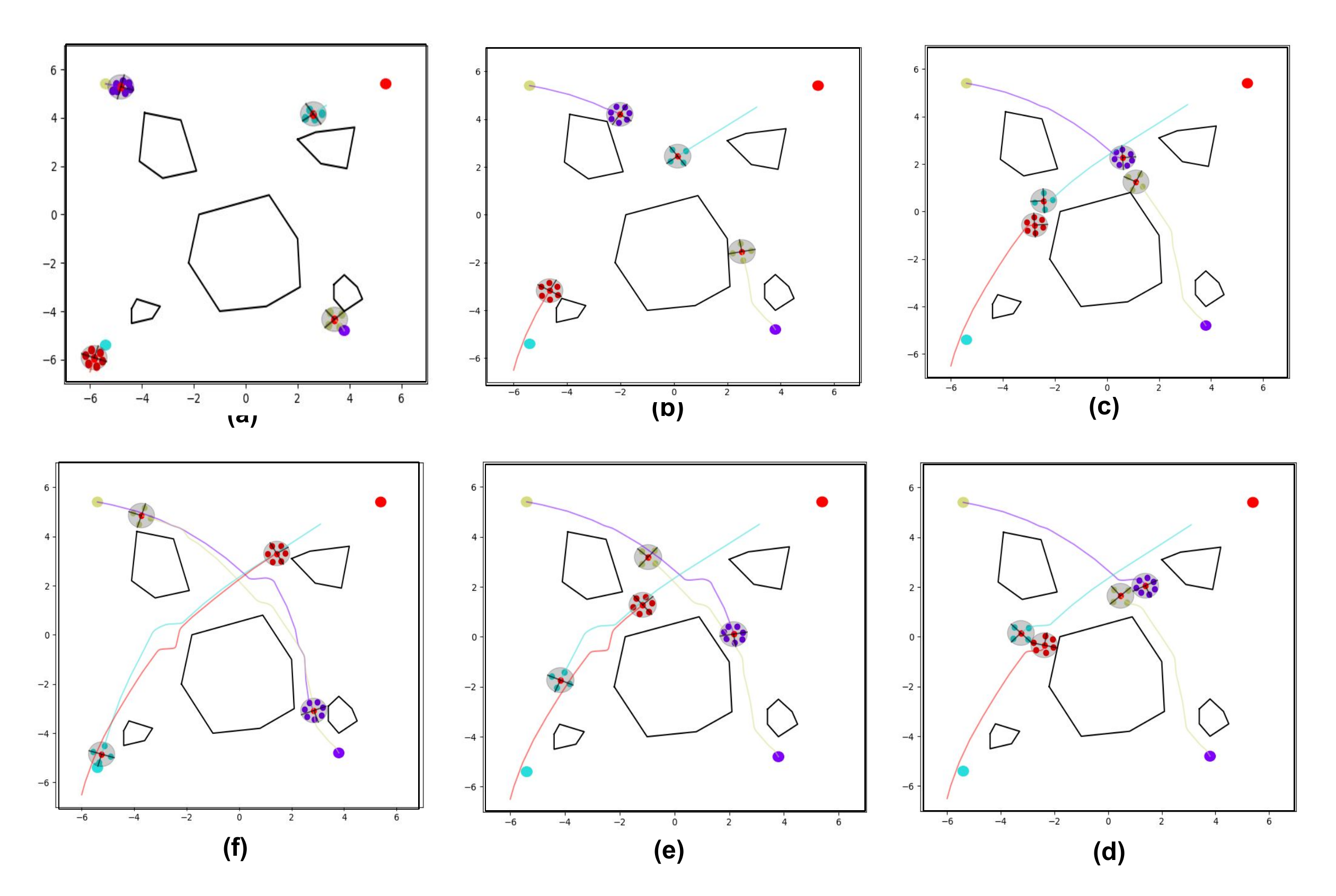}
	\caption{Four PTS are running avoiding collision with static obstacles and other transport systems}
	\label{static_obs_setup}
\end{figure}
We consider four payload transport systems carrying a payload. Five static obstacles are considered at random places in the arena. Each payload transport system receives the trajectory information using Algorithm \ref{alg:algo1} such that there is no collision amongst the PTS and static obstacles. Fig. \ref{static_obs_setup} shows a initial set up of all formations and the obstacles present in the surroundings. We observe that all the systems successfully avoid collision with the obstacles. \par
\begin{figure}[h]
    \centering
	\includegraphics[width=\linewidth, height=4.5cm]{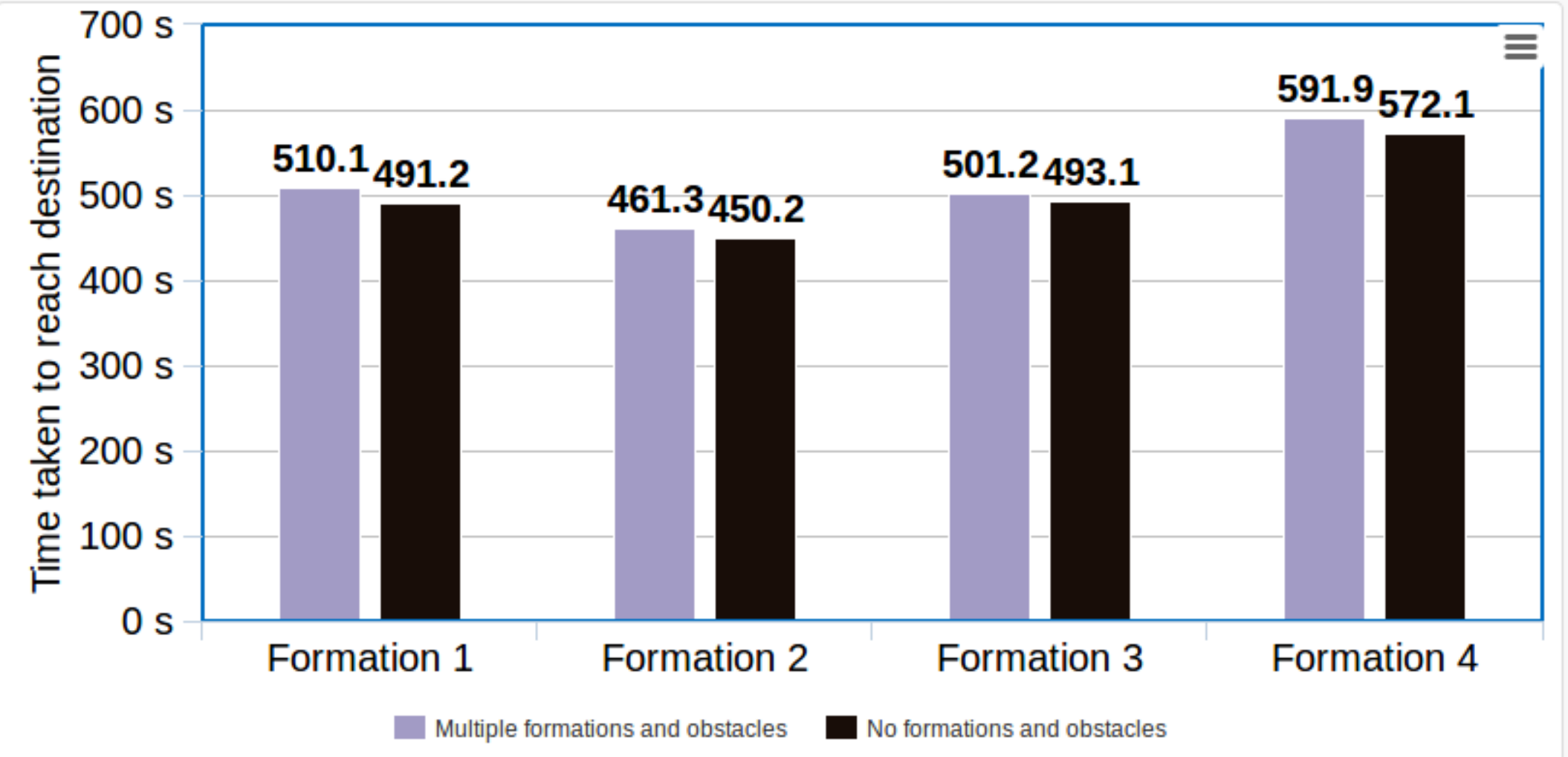}
	\caption{Comparison of time taken by our system with and without any traffic and static obstacles in the environment}
	\label{static_obs_time}
\end{figure} 
We compare the time taken by the PTS to reach to their respective destinations with and without obstacles. Fig. \ref{static_obs_time} shows that the time taken by the PTS in presence of obstacles is comparable to the time taken by the PTS in no obstacles scenario. We conclude that an optimal collision avoiding path is provided by our control algorithm. \par
\begin{figure}[h]
    \centering
	\includegraphics[width=\linewidth, height=5cm]{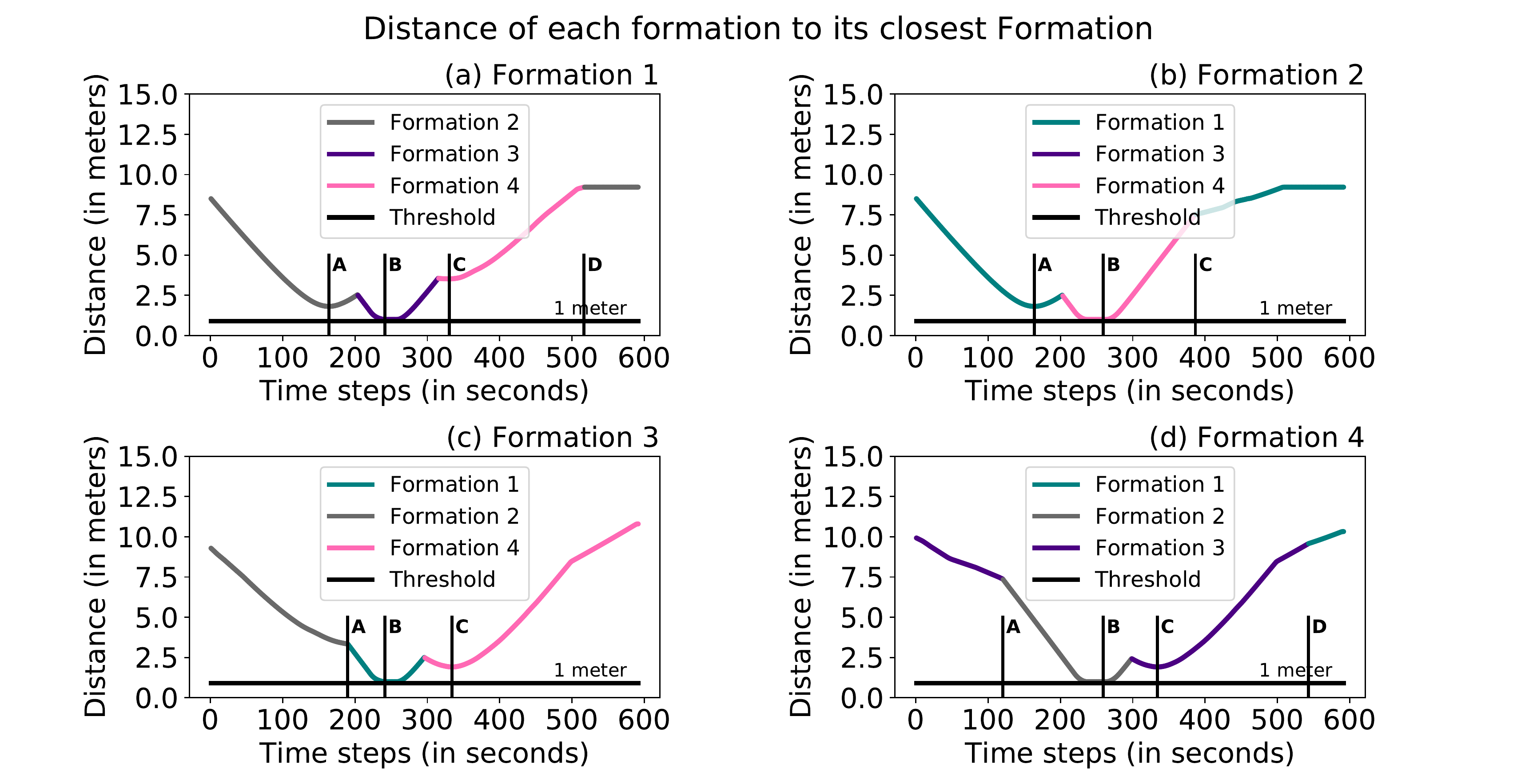}
	\caption{Distance of each PTS with its closest neighbourhood in presence of static obstacles.}
	\label{static_obs_dist}
\end{figure}
\begin{figure}[h]
    \centering
	\includegraphics[width=\linewidth, height=5cm]{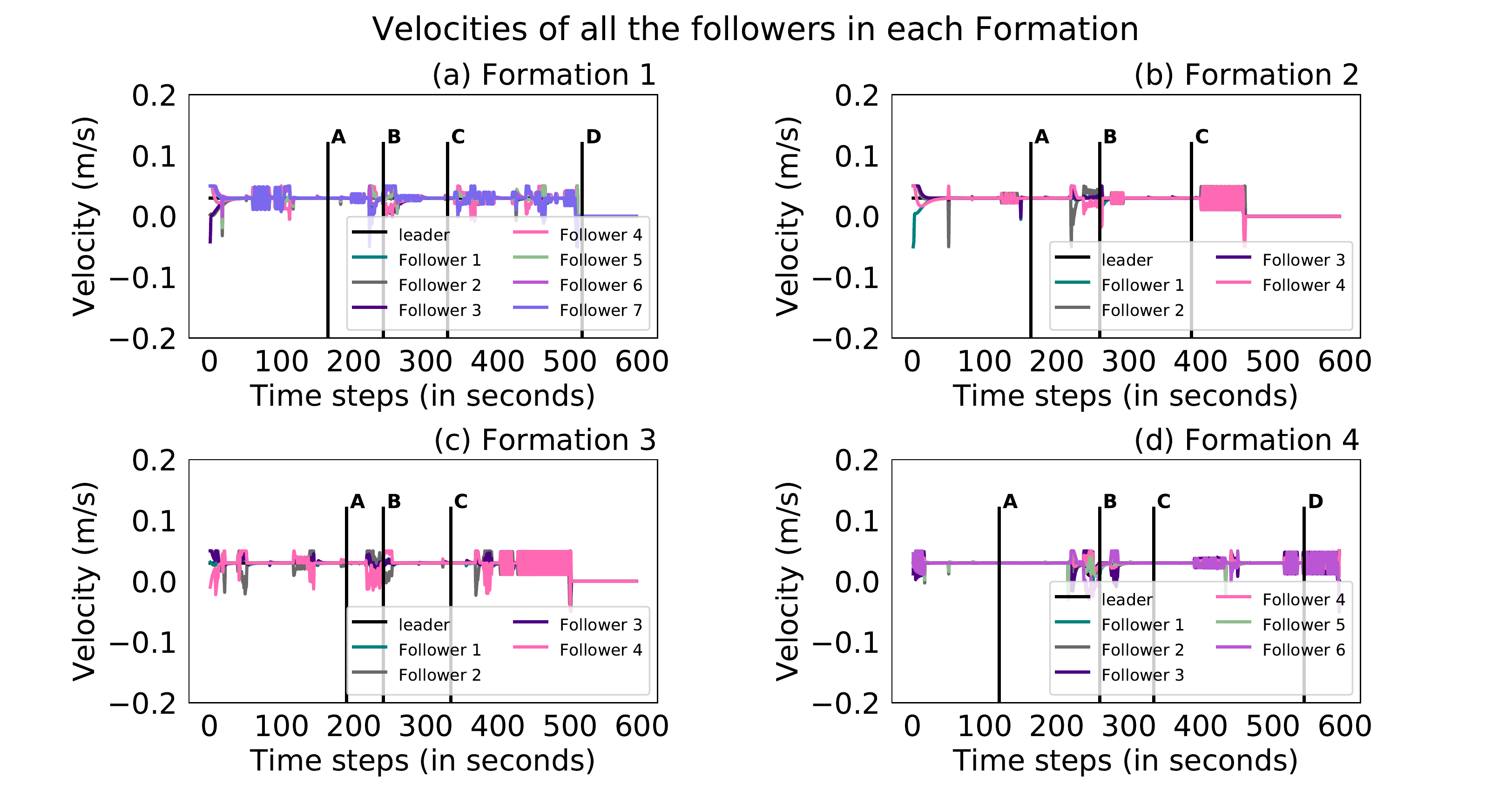}
	\caption{Robots velocities of all PTS with static obstacles.}
	\label{static_obs_vel}
\end{figure}

In this case, we plot the minimum distance of a PTS which is to be maintained wrt. neighbourhood PTS (similar to the previous case with no static obstacle). This will make sure that the formations do not collide with each other. Fig. \ref{static_obs_dist} shows the minimum distance maintained by each PTS with its closest neighbourhood PTS. We present a change in the velocities of each PTS when coming close to each other Fig. \ref{static_obs_vel}. We observe, a PTS changes its velocities (as generated by the nh-ORCA) such that it avoids any collision with other obstacles. As discussed in the above case, we mark points in Fig. \ref{static_obs_dist} and Fig. \ref{static_obs_vel} to relate the change in velocities of the formation in the vicinity of the other formations or static obstacles. For example, formation $2$ and formation $4$ come close to each other at Point B and the velocity of both the formation also changes at the same to avoid any obstacle (see Fig. \ref{static_obs_vel}). \par
\begin{figure}[h]
    \centering
	\includegraphics[width=\linewidth, height=5cm]{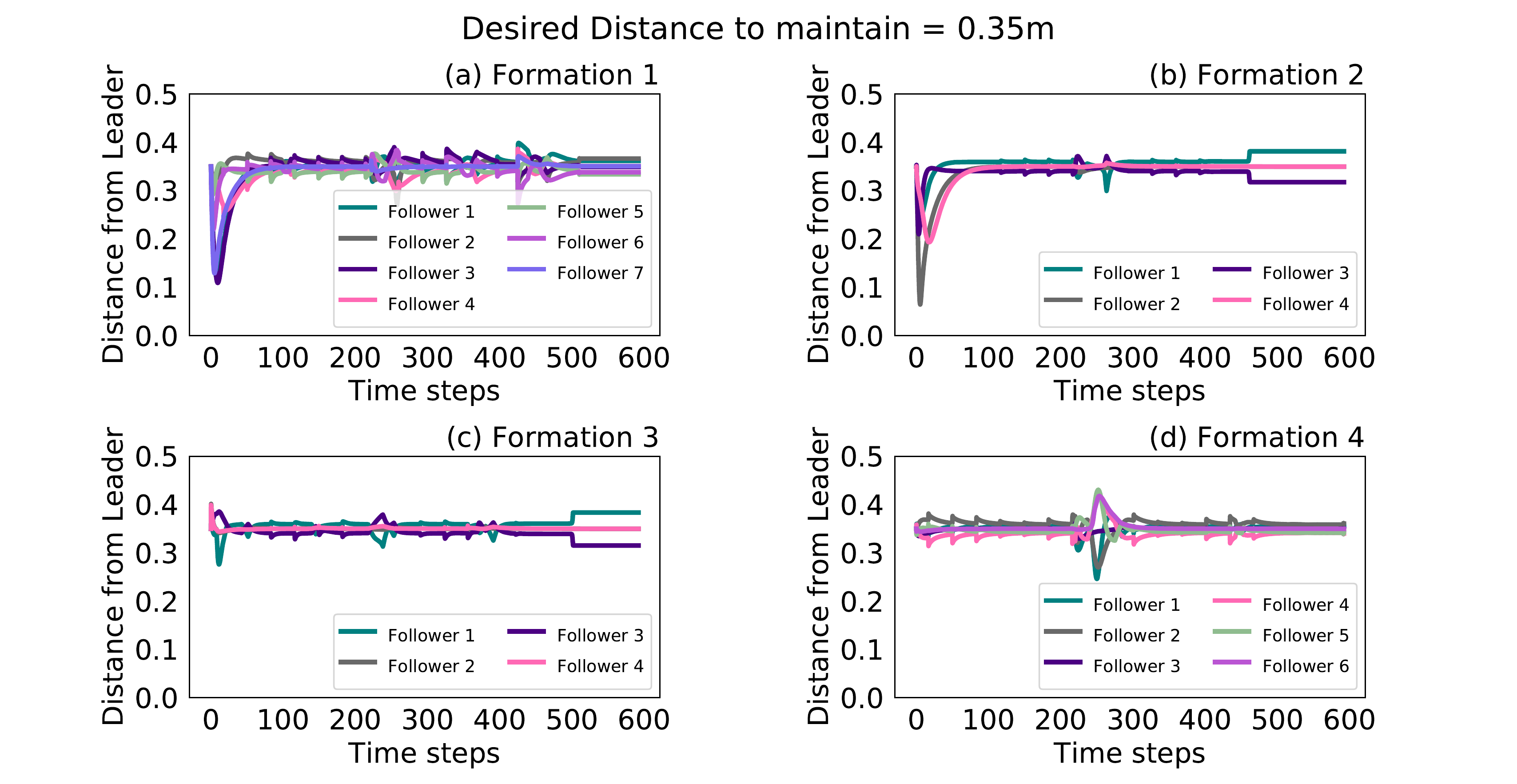}
	\caption{Distance of all the robots in each formation in presence of static obstacles}
	\label{static_obs_fd}
\end{figure}
It is even more challenging in this case to maintain the formation for each PTS as the static obstacles in the surroundings make it more difficult to maintain the desired distance and angles of the follower with the leader robot. Fig. \ref{static_obs_fd} highlights that the followers in each PTS is maintaining the specified parameters with the leader when avoiding the obstacles. For simulation of this scenario also, we have taken the distance of followers w.r.t to leader to be $0.35 m$. 
\subsection{Scalability analysis of the system}
\begin{figure}[h]
    \centering
	\includegraphics[width=\linewidth, height=4.5cm]{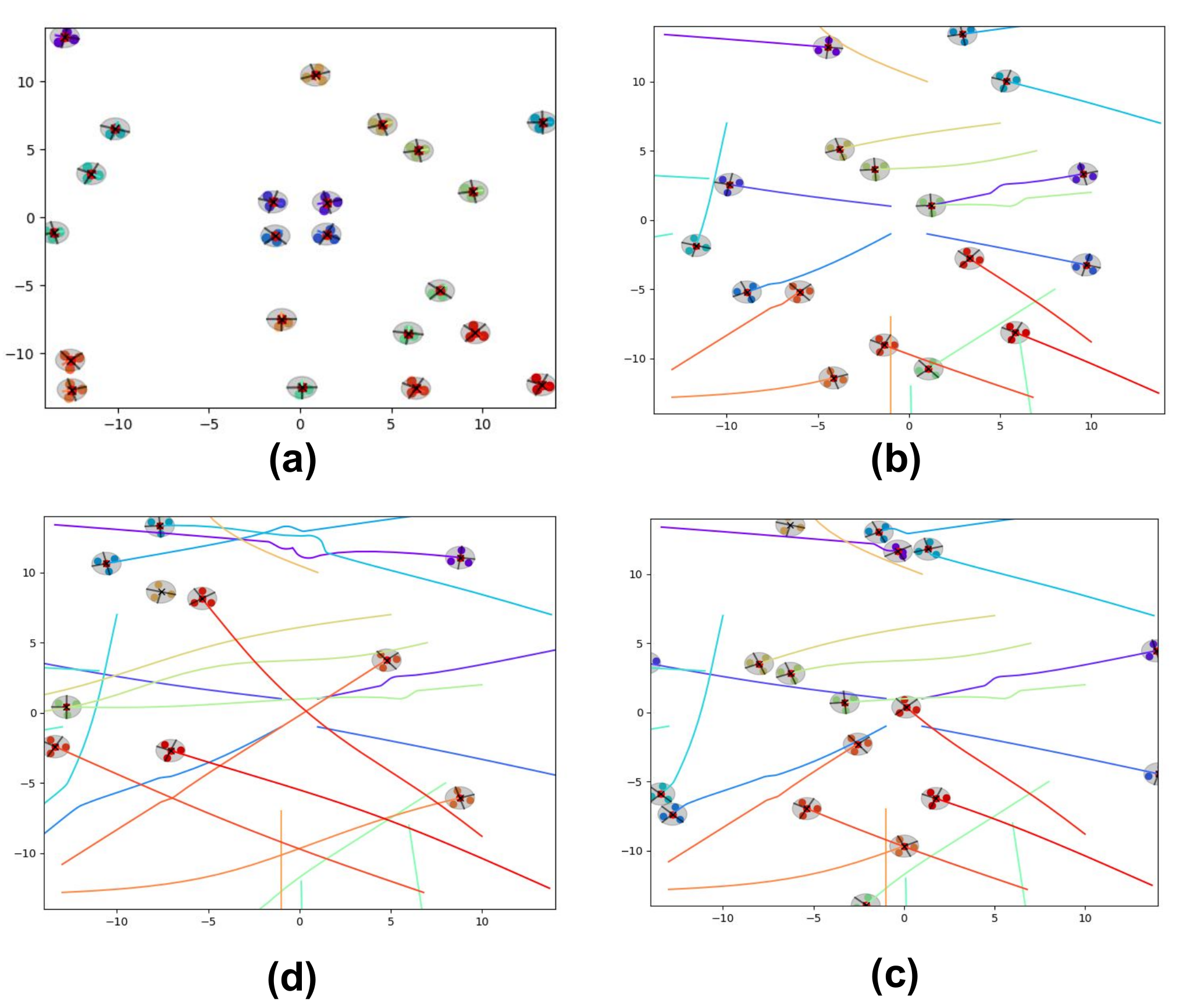}
	\caption{Thirty different PTS are moving to carry a payload to their respective destination without any collision.}
	\label{scalable}
\end{figure}

Our system scales well with the increase in number of PTS, hence making it more efficient and robust. We showcase through simulations, that our system works well with $30$ different payload transport systems with variable number of robots in each of them. But, in general, it can be extended to any number of PTS. Fig. \ref{scalable} shows that all the PTS reach their destination to complete a given task without colliding with any of the neighbourhood PTS. Thus our system is more robust and reliable to work with large number of PTS in an industrial/warehouse environment. We have added more detailed results and simulations of the scalability in the video uploaded.
\vspace{-0.2cm}
\section{Conclusion}
\label{conclusion}
We simulate the traffic management strategies for multiple loosely coupled payload transport system. Each PTS comprises of a number of non-holonomic mobile robots which are moving in a formation and carrying a payload from one place to another. The formation control is done using a decentralized leader-follower based algorithm. A modified version of decentralized collision avoidance algorithm nh-ORCA is used to compute the collision-free velocities for the leader of the formation. The followers follow the leader such that a rigid formation is maintained while simultaneously moving towards a goal. Various environments are considered for the simulation which includes \begin{enumerate*}[label=(a),font=\itshape]
    \item Multiple PTS are running with no other obstacles
\end{enumerate*}  
\begin{enumerate*}[label=(b),font=\itshape]
    \item Multiple PTS are running with static obstacles
\end{enumerate*}. A comparison is made with a baseline scenario where individual PTS are moving to reach a goal.
\bibliographystyle{ieeetr}
\bibliography{root}

\end{document}